\documentclass[preprintnumbers,amsmath,amssymb,showpacs,twocolomns]{revtex4}

\usepackage{graphicx}
\usepackage{dcolumn}
\usepackage{bm}

\begin{document}

\title{ Probabilistic teleportation of unknown two-particle state  via  POVM
}

\author{ Fengli Yan $^{1,2}$,  Hewei Ding$^1$}

\affiliation {$^1$  College of Physics Science and Information Engineering, Hebei Normal University, Shijiazhuang 050016, China\\
 $^2$ CCAST (World Laboratory), P.O. Box 8730, Beijing 100080, China }

\date{\today}

\begin{abstract}
We propose a scheme for probabilistic teleportation of unknown two-particle state with partly entangled
four-particle state via POVM. In this scheme the teleportation of unknown two-particle state  can be realized
with certain probability by performing two Bell state measurements, a proper POVM  and a unitary transformation.
\end{abstract}


\maketitle

Quantum teleportation is a process of transmission of an unknown quantum state via a previously shared EPR pair
with the help of only two classical bits transmitted through a classical channel \cite {Bennett93}. It was
regarded as one of the most striking
 progress of quantum information theory \cite {NielsenChuang}. It may  have a number of useful
 applications in quantum computer \cite {CiracZoller,Barenco},
quantum dense coding \cite {BennenttWiesner}, quantum cryptography \cite {BB84,Ekert,Bennett92} and quantum
secure direct communication \cite {ShimizuImoto,Beige,DengLong,YanZhang,Gao}.

Since  Bennett et al.  showed that an unknown quantum state of two-state particle (or qubit) can be teleported
from a sender Alice to a spatially distant receiver Bob in 1993 \cite{Bennett93}, people have paid much
attention to quantum teleportation. Research work on quantum teleportation was soon widely started  up, and has
got great development, theoretical and experimental as well.  On the one hand, the teleportation of a photon
polarization state has been demonstrated experimentally with the use of polarization-entangled photons \cite
{BPZ} and path-entangled photons \cite {Boschi}, respectively.  The teleportation  of a coherent state
corresponding to continuous variable system was also  realized in the laboratory \cite {Furusawa}. On the other
hand, the idea of quantum teleportation has been generalized  to many cases \cite {GorbachevTrubilko,
LuGuo,Bandyopadhyay, Rigolin, ShiGuo, ZengLong, MorHorodecki, LiuGuo, DaiLi, YanTanYang, YanWang, YanYang,
YanBai, GaoYanWang, GaoWangYanChinPhysLett, GaoYanWangChinesePhysics, Gao2}.

 Mor and Horodecki discussed the problems of teleportation via positive operator valued measure (POVM) \cite {Helstrom}, which was also
  called generalized measurement, and conclusive teleportation \cite{MorHorodecki}.
 It was showed that a perfect conclusive teleportation  can be obtained with any pure entangled state.
Bandyopadhyay   presented two optimal methods of teleporting an unknown qubit using any pure entangled state
\cite{Bandyopadhyay}.  About three years ago we designed  a  scheme for probabilistic teleporting an unknown
two-particle state  with partly pure entangled four-particle state via a projective measurement on an auxiliary
particle \cite {YanTanYang}. Is it possible to use POVM  in the probabilistic teleportation of  an unknown
two-particle state  with partly pure entangled four-particle state? In this Letter, we will try to answer this
question, i.e., to propose  a scheme for probabilistic teleportation of unknown two-particle state  with partly
entangled four-particle state via  POVM. It  will be shown that  by performing two Bell state measurements, a
proper POVM  and a unitary transformation, the unknown two-particle state  can be teleported from the sender
Alice to the receiver Bob  with certain probability. We  state our scheme in details as follows.

Suppose that the sender Alice has two particles 1,2 in an unknown state
\begin{equation}
|\Phi\rangle_{12}=(a|00\rangle+b|01\rangle+c|10\rangle+d|11\rangle)_{12},
\end{equation}
where $a, b, c, d$ are arbitrary complex numbers, and satisfy $|a|^2+|b|^2+|c|^2+|d|^2=1$. We also suppose that
Alice and Bob share  quantum entanglement in the form of  following  partly pure entangled four-particle state,
which will be used as the quantum channel,
\begin{equation}
|\Phi\rangle_{3456}=(\alpha|0000\rangle+\beta|1001\rangle+\gamma|0110\rangle+\delta|1111\rangle)_{3456},
\end{equation}
where $\alpha, \beta, \gamma, \delta$ are nonzero real numbers, and $\alpha^2+\beta^2+\gamma^2+\delta^2=1$.  The
particles 3 and 4, and particle pair (1,2) are  in Alice's possession, and other two particles 5 and 6 are in
Bob's possession. The overall state of six particles is
\begin{equation}
|\Phi\rangle_w=|\Phi\rangle_{12}\otimes|\Phi\rangle_{3456}.
\end{equation}
In order to realize the teleportation, firstly Alice performs  two Bell state measurements on particles 2,3 and
1,4, then the  resulting state of Bob's particles 5,6 will be one of the following states \cite{YanTanYang}
\begin{equation}
|\Psi_0\rangle_{56}=\frac
{_{14}\langle\Phi^+|_{23}\langle\Phi^+|\Phi\rangle_w}{|_{14}\langle\Phi^+|_{23}\langle\Phi^+|\Phi\rangle_w|}=\frac
{1}{\sqrt {|a\alpha|^2+|b\beta|^2+|c\gamma|^2+|d\delta|^2}}(a\alpha|00\rangle+b\beta|01\rangle+
c\gamma|10\rangle+d\delta|11\rangle)_{56},
\end{equation}
\begin{equation}
|\Psi_1\rangle_{56}=\frac
{_{14}\langle\Phi^-|_{23}\langle\Phi^+|\Phi\rangle_w}{|_{14}\langle\Phi^-|_{23}\langle\Phi^+|\Phi\rangle_w|}=\frac
{1}{\sqrt {|a\alpha|^2+|b\beta|^2+|c\gamma|^2+|d\delta|^2}}(a\alpha|00\rangle+b\beta|01\rangle
-c\gamma|10\rangle-d\delta|11\rangle)_{56},
\end{equation}
\begin{equation}
|\Psi_2\rangle_{56}=\frac
{_{14}\langle\Psi^+|_{23}\langle\Phi^+|\Phi\rangle_w}{|_{14}\langle\Psi^+|_{23}\langle\Phi^+|\Phi\rangle_w|}=\frac
{1}{\sqrt {|a\gamma|^2+|b\delta|^2+|c\alpha|^2+|d\beta|^2}}(a\gamma|10\rangle+b\delta|11\rangle
+c\alpha|00\rangle+d\beta|01\rangle)_{56},
\end{equation}
\begin{equation}
|\Psi_3\rangle_{56}=\frac
{_{14}\langle\Psi^+|_{23}\langle\Phi^-|\Phi\rangle_w}{|_{14}\langle\Psi^+|_{23}\langle\Phi^-|\Phi\rangle_w|}=\frac
{1}{\sqrt {|a\gamma|^2+|b\delta|^2+|c\alpha|^2+|d\beta|^2}}(a\gamma|10\rangle+b\delta|11\rangle
-c\alpha|00\rangle-d\beta|01\rangle)_{56},
\end{equation}
\begin{equation}
|\Psi_4\rangle_{56}=\frac
{_{14}\langle\Phi^+|_{23}\langle\Phi^-|\Phi\rangle_w}{|_{14}\langle\Phi^+|_{23}\langle\Phi^-|\Phi\rangle_w|}=\frac
{1}{\sqrt {|a\alpha|^2+|b\beta|^2+|c\gamma|^2+|d\delta|^2}}(a\alpha|00\rangle-b\beta|01\rangle
+c\gamma|10\rangle-d\delta|11\rangle)_{56},
\end{equation}
\begin{equation}
|\Psi_5\rangle_{56}=\frac
{_{14}\langle\Phi^-|_{23}\langle\Phi^-|\Phi\rangle_w}{|_{14}\langle\Phi^-|_{23}\langle\Phi^-|\Phi\rangle_w|}=\frac
{1}{\sqrt {|a\alpha|^2+|b\beta|^2+|c\gamma|^2+|d\delta|^2}}(a\alpha|00\rangle-b\beta|01\rangle
-c\gamma|10\rangle+d\delta|11\rangle)_{56},
\end{equation}
\begin{equation}
|\Psi_6\rangle_{56}=\frac
{_{14}\langle\Psi^+|_{23}\langle\Phi^-|\Phi\rangle_w}{|_{14}\langle\Psi^+|_{23}\langle\Phi^-|\Phi\rangle_w|}=\frac
{1}{\sqrt {|a\gamma|^2+|b\delta|^2+|c\alpha|^2+|d\beta|^2}}(a\gamma|10\rangle-b\delta|11\rangle
+c\alpha|00\rangle-d\beta|01\rangle)_{56},
\end{equation}
\begin{equation}
|\Psi_7\rangle_{56}=\frac
{_{14}\langle\Psi^-|_{23}\langle\Phi^-|\Phi\rangle_w}{|_{14}\langle\Psi^-|_{23}\langle\Phi^-|\Phi\rangle_w|}=\frac
{1}{\sqrt {|a\gamma|^2+|b\delta|^2+|c\alpha|^2+|d\beta|^2}}(a\gamma|10\rangle-b\delta|11\rangle
-c\alpha|00\rangle+d\beta|01\rangle)_{56},
\end{equation}
\begin{equation}
|\Psi_8\rangle_{56}=\frac
{_{14}\langle\Phi^+|_{23}\langle\Psi^+|\Phi\rangle_w}{|_{14}\langle\Phi^+|_{23}\langle\Psi^+|\Phi\rangle_w|}=\frac
{1}{\sqrt {|a\beta|^2+|b\alpha|^2+|c\delta|^2+|d\gamma|^2}}(a\beta|01\rangle+b\alpha|00\rangle
+c\delta|11\rangle+d\gamma|10\rangle)_{56},
\end{equation}
\begin{equation}
|\Psi_9\rangle_{56}=\frac
{_{14}\langle\Phi^-|_{23}\langle\Psi^+|\Phi\rangle_w}{|_{14}\langle\Phi^-|_{23}\langle\Psi^+|\Phi\rangle_w|}=\frac
{1}{\sqrt {|a\beta|^2+|b\alpha|^2+|c\delta|^2+|d\gamma|^2}}(a\beta|01\rangle+b\alpha|00\rangle
-c\delta|11\rangle-d\gamma|10\rangle)_{56},
\end{equation}
\begin{equation}
|\Psi_{10}\rangle_{56}=\frac
{_{14}\langle\Psi^+|_{23}\langle\Psi^+|\Phi\rangle_w}{|_{14}\langle\Psi^+|_{23}\langle\Psi^+|\Phi\rangle_w|}=\frac
{1}{\sqrt {|a\delta|^2+|b\gamma|^2+|c\beta|^2+|d\alpha|^2}}(a\delta|11\rangle+b\gamma|10\rangle
+c\beta|01\rangle+d\alpha|00\rangle)_{56},
\end{equation}
\begin{equation}
|\Psi_{11}\rangle_{56}=\frac
{_{14}\langle\Psi^-|_{23}\langle\Psi^+|\Phi\rangle_w}{|_{14}\langle\Psi^-|_{23}\langle\Psi^+|\Phi\rangle_w|}=\frac
{1}{\sqrt {|a\delta|^2+|b\gamma|^2+|c\beta|^2+|d\alpha|^2}}(a\delta|11\rangle+b\gamma|10\rangle
-c\beta|01\rangle-d\alpha|00\rangle)_{56},
\end{equation}
\begin{equation}
|\Psi_{12}\rangle_{56}=\frac
{_{14}\langle\Phi^+|_{23}\langle\Psi^-|\Phi\rangle_w}{|_{14}\langle\Phi^+|_{23}\langle\Psi^-|\Phi\rangle_w|}=\frac
{1}{\sqrt {|a\beta|^2+|b\alpha|^2+|c\delta|^2+|d\gamma|^2}}(a\beta|01\rangle-b\alpha|00\rangle
+c\delta|11\rangle-d\gamma|10\rangle)_{56},
\end{equation}
\begin{equation}
|\Psi_{13}\rangle_{56}=\frac
{_{14}\langle\Phi^-|_{23}\langle\Psi^-|\Phi\rangle_w}{|_{14}\langle\Phi^-|_{23}\langle\Psi^-|\Phi\rangle_w|}=\frac
{1}{\sqrt {|a\beta|^2+|b\alpha|^2+|c\delta|^2+|d\gamma|^2}}(a\beta|01\rangle-b\alpha|00\rangle
-c\delta|11\rangle+d\gamma|10\rangle)_{56},
\end{equation}
\begin{equation}
|\Psi_{14}\rangle_{56}=\frac
{_{14}\langle\Psi^+|_{23}\langle\Psi^-|\Phi\rangle_w}{|_{14}\langle\Psi^+|_{23}\langle\Psi^-|\Phi\rangle_w|}=\frac
{1}{\sqrt {|a\delta|^2+|b\gamma|^2+|c\beta|^2+|d\alpha|^2}}(a\delta|11\rangle-b\gamma|10\rangle
+c\beta|01\rangle-d\alpha|00\rangle)_{56},
\end{equation}
\begin{equation}
|\Psi_{15}\rangle_{56}=\frac
{_{14}\langle\Psi^-|_{23}\langle\Psi^-|\Phi\rangle_w}{|_{14}\langle\Psi^-|_{23}\langle\Psi^-|\Phi\rangle_w|}=\frac
{1}{\sqrt {|a\delta|^2+|b\gamma|^2+|c\beta|^2+|d\alpha|^2}}(a\delta|11\rangle-b\gamma|10\rangle
-c\beta|01\rangle+d\alpha|00\rangle)_{56}.
\end{equation}
Here
\begin{equation}\begin{array}{ccc}
|\Phi^\pm\rangle=\frac {1}{\sqrt 2}(|00\rangle\pm|11\rangle),&~& |\Psi^\pm\rangle=\frac {1}{\sqrt
2}(|01\rangle\pm|10\rangle) \end{array}
\end{equation}
are the Bell states. Then Alice informs Bob her two Bell state measurement outcomes via a classical channel. By
 outcomes received, Bob can determine the state of particles 5,6 exactly. Without loss of generality, next we
will give  the case for  $|\Psi_0\rangle_{56}$, the  other cases  can be deduced similarly. In order to realize
the teleportation, Bob  introduces two auxiliary qubits $a, b$ in the  state $|00\rangle_{ab}$. So the state of
particles $5,6,a,b$ becomes
\begin{equation}
|\Psi_0\rangle_{56}|00\rangle_{ab}=\frac {1}{\sqrt
{|a\alpha|^2+|b\beta|^2+|c\gamma|^2+|d\delta|^2}}(a\alpha|00\rangle+b\beta|01\rangle+
c\gamma|10\rangle+d\delta|11\rangle)_{56}|00\rangle_{ab}.
\end{equation}
Then Bob performs two controlled-not operations with particles 5,6 as the controlled qubits and the auxiliary
particles $a,b$ as the target qubits respectively. After completing this operation the particles $5,6,a,b$ are
in state
\begin{equation}
|\Psi'_0\rangle_{56ab}=\frac {1}{\sqrt
{|a\alpha|^2+|b\beta|^2+|c\gamma|^2+|d\delta|^2}}(a\alpha|0000\rangle+b\beta|0101\rangle+
c\gamma|1010\rangle+d\delta|1111\rangle)_{56ab}.
\end{equation}
 A simple algebraic rearrangement of this expression
yields
\begin{equation}
\begin{array}{cl}
|\Psi'_0\rangle_{56ab}=&\frac {1}{4\sqrt
{|a\alpha|^2+|b\beta|^2+|c\gamma|^2+|d\delta|^2}}[(a|00\rangle+b|01\rangle+c|10\rangle+d|11\rangle)_{56}\otimes(\alpha|00\rangle+\beta|01\rangle
+\gamma|10\rangle+\delta|11\rangle)_{ab}\\
~&+(a|00\rangle+b|01\rangle-c|10\rangle-d|11\rangle)_{56}\otimes(\alpha|00\rangle+\beta|01\rangle
-\gamma|10\rangle-\delta|11\rangle)_{ab}\\
~&+(a|00\rangle-b|01\rangle+c|10\rangle-d|11\rangle)_{56}\otimes(\alpha|00\rangle-\beta|01\rangle
+\gamma|10\rangle-\delta|11\rangle)_{ab}\\
~&+(a|00\rangle-b|01\rangle-c|10\rangle+d|11\rangle)_{56}\otimes(\alpha|00\rangle-\beta|01\rangle
-\gamma|10\rangle+\delta|11\rangle)_{ab}].
\end{array}
\end{equation}
At this stage, Bob makes an optimal POVM  \cite {Bandyopadhyay,Helstrom} on the ancillary particle $a, b$ to
conclusively distinguish the above states. We choose the optimal POVM in this subspace as follows
\begin{equation}
\begin{array}{ccccc}
P_1=\frac {1}{x}|\Psi_1\rangle\langle\Psi_1|,& P_2=\frac {1}{x}|\Psi_2\rangle\langle\Psi_2|,& P_3=\frac
{1}{x}|\Psi_3\rangle\langle\Psi_3|, & P_4=\frac {1}{x}|\Psi_4\rangle\langle\Psi_4|, & P_5=I-\frac
{1}{x}\Sigma_{i=1}^4|\Psi_i\rangle\langle\Psi_i|,
\end{array} \end{equation}
where
\begin{equation}
 |\Psi_1\rangle=\frac {1}{\sqrt {\frac {1}{\alpha^2}+\frac
{1}{\beta^2}+\frac {1}{\gamma^2}+\frac {1}{\delta ^2}}}(\frac {1}{\alpha}|00\rangle+\frac
{1}{\beta}|01\rangle+\frac {1}{\gamma}|10\rangle+\frac {1}{\delta}|11\rangle)_{ab},
\end{equation}
\begin{equation}
 |\Psi_2\rangle=\frac {1}{\sqrt {\frac {1}{\alpha^2}+\frac
{1}{\beta^2}+\frac {1}{\gamma^2}+\frac {1}{\delta ^2}}}(\frac {1}{\alpha}|00\rangle+\frac
{1}{\beta}|01\rangle-\frac {1}{\gamma}|10\rangle-\frac {1}{\delta}|11\rangle)_{ab},
\end{equation}
\begin{equation}
 |\Psi_3\rangle=\frac {1}{\sqrt {\frac {1}{\alpha^2}+\frac
{1}{\beta^2}+\frac {1}{\gamma^2}+\frac {1}{\delta ^2}}}(\frac {1}{\alpha}|00\rangle-\frac
{1}{\beta}|01\rangle+\frac {1}{\gamma}|10\rangle-\frac {1}{\delta}|11\rangle)_{ab},
\end{equation}
\begin{equation}
 |\Psi_4\rangle=\frac {1}{\sqrt {\frac {1}{\alpha^2}+\frac
{1}{\beta^2}+\frac {1}{\gamma^2}+\frac {1}{\delta ^2}}}(\frac {1}{\alpha}|00\rangle-\frac
{1}{\beta}|01\rangle-\frac {1}{\gamma}|10\rangle+\frac {1}{\delta}|11\rangle)_{ab};
\end{equation}
  $I$ is an identity operator;  $x$ is  a coefficient  relating  to $\alpha, \beta, \gamma, \delta$,
 $1\leq x\leq 4$, and  makes  $P_5$ to be  a positive operator. For  exactly determining $x$, we
would like to write the five  operators ${P_1, P_2, P_3,P_4, P_5}$ in the matrix form
\begin{equation}
P_1=\frac {1}{x(\frac {1}{\alpha^2}+\frac {1}{\beta^2}+\frac {1}{\gamma^2}+\frac {1}{\delta ^2})}\left
(\begin{array}{cccc} \frac {1}{\alpha^2}& \frac {1}{\alpha\beta} & \frac {1} {\alpha\gamma} & \frac
{1}{\alpha\delta}\\
\frac {1}{\alpha\beta}& \frac {1}{\beta^2} & \frac {1} {\beta\gamma} & \frac
{1}{\beta\delta}\\
\frac {1}{\alpha\gamma}& \frac {1}{\beta\gamma} & \frac {1} {\gamma^2} & \frac
{1}{\gamma\delta}\\
\frac {1}{\alpha\delta}& \frac {1}{\beta\delta} & \frac {1} {\gamma\delta} & \frac
{1}{\delta^2}\\
\end{array}\right ),
\end{equation}
\begin{equation}
P_2=\frac {1}{x(\frac {1}{\alpha^2}+\frac {1}{\beta^2}+\frac {1}{\gamma^2}+\frac {1}{\delta ^2})}\left
(\begin{array}{cccc} \frac {1}{\alpha^2}& \frac {1}{\alpha\beta} & -\frac {1} {\alpha\gamma} & -\frac
{1}{\alpha\delta}\\
\frac {1}{\alpha\beta}& \frac {1}{\beta^2} & -\frac {1} {\beta\gamma} & -\frac
{1}{\beta\delta}\\
-\frac {1}{\alpha\gamma} & -\frac {1}{\beta\gamma} & \frac {1} {\gamma^2} & \frac
{1}{\gamma\delta}\\
-\frac {1}{\alpha\delta} & -\frac {1}{\beta\delta} & \frac {1} {\gamma\delta} & \frac
{1}{\delta^2}\\
\end{array}\right ),
\end{equation}
\begin{equation}
P_3=\frac {1}{x(\frac {1}{\alpha^2}+\frac {1}{\beta^2}+\frac {1}{\gamma^2}+\frac {1}{\delta ^2})}\left
(\begin{array}{cccc} \frac {1}{\alpha^2}& -\frac {1}{\alpha\beta} & \frac {1} {\alpha\gamma} & -\frac
{1}{\alpha\delta}\\
-\frac {1}{\alpha\beta}& \frac {1}{\beta^2} & -\frac {1} {\beta\gamma} & \frac
{1}{\beta\delta}\\
\frac {1}{\alpha\gamma}& -\frac {1}{\beta\gamma} & \frac {1} {\gamma^2} & -\frac
{1}{\gamma\delta}\\
-\frac {1}{\alpha\delta}& \frac {1}{\beta\delta} & -\frac {1} {\gamma\delta} & \frac
{1}{\delta^2}\\
\end{array}\right ),
\end{equation}
\begin{equation}
P_4=\frac {1}{x(\frac {1}{\alpha^2}+\frac {1}{\beta^2}+\frac {1}{\gamma^2}+\frac {1}{\delta ^2})}\left
(\begin{array}{cccc} \frac {1}{\alpha^2}& -\frac {1}{\alpha\beta} & -\frac {1} {\alpha\gamma} & \frac
{1}{\alpha\delta}\\
-\frac {1}{\alpha\beta}& \frac {1}{\beta^2} & \frac {1} {\beta\gamma} & -\frac
{1}{\beta\delta}\\
-\frac {1}{\alpha\gamma}& \frac {1}{\beta\gamma} & \frac {1} {\gamma^2} & -\frac
{1}{\gamma\delta}\\
\frac {1}{\alpha\delta}& -\frac {1}{\beta\delta} & -\frac {1} {\gamma\delta} & \frac
{1}{\delta^2}\\
\end{array}\right ),
\end{equation}
\begin{equation}\begin{array}{l}
P_5=\frac {1}{\frac {1}{\alpha^2}+\frac {1}{\beta^2}+\frac {1}{\gamma^2}+\frac {1}{\delta ^2}}\\
\left
(\begin{array}{cccc} (1-\frac {4}{x})\frac {1}{\alpha^2}+\frac {1}{\beta^2}+\frac {1}{\gamma^2}+\frac
{1}{\delta^2} & 0 & 0 & 0\\
0 & (1-\frac {4}{x})\frac {1}{\beta^2}+\frac {1}{\alpha^2}+\frac {1}{\gamma^2}+\frac {1}{\delta^2} & 0 & 0\\
0 & 0 & (1-\frac {4}{x})\frac {1}{\gamma^2}+\frac {1}{\alpha^2}+\frac {1}{\beta^2}+\frac {1}{\delta^2} & 0\\
0 & 0 & 0 & (1-\frac {4}{x})\frac {1}{\delta^2}+\frac {1}{\alpha^2}+\frac {1}{\beta^2}+\frac {1}{\gamma^2}\\
\end{array}\right ).
\end{array}
\end{equation}
Obviously, we should carefully choose $x$ such  that  all the diagonal elements of $P_5$ are  nonnegative. If
the result of Bob's POVM  is $P_1$, then Bob can safely conclude that the state of the particles 5,6 is
\begin{equation}\label{telestate}
|\Phi\rangle_{56}=(a|00\rangle+b|01\rangle+c|10\rangle+d|11\rangle)_{56}.
\end{equation}
  If Bob's POVM outcome is $P_2$, Bob can obtain
$|\Phi\rangle_{56}$ by applying the unitary transformation $\sigma_z\otimes I$ on the particles 5,6. If the
Bob's POVM outcome $P_3$ occurs, Bob makes $I\otimes \sigma_z$ on the  particles 5,6 to recover
$|\Phi\rangle_{56}$. If Bob's POVM outcome is  $P_4$, Bob applies   $\sigma_z\otimes \sigma_z$ on the particles
5,6 to recover $|\Phi\rangle_{56}$. Therefore, in these four cases stated above, the teleportation is realized
successfully. However if Bob's POVM outcome is $P_5$, Bob can infer nothing about the identity of the state of
the particles 5,6. In this case the teleportation fails. As a matter of fact, the key point is that Bob never
makes a mistake identifying the state of the particles 5,6. This infallibility comes at the price that sometime
Bob obtains no information about the identity of the state of the particles 5,6.

By the similar method we can make the teleportation successful in the other outcomes of Alice's  Bell state
measurement. For the sake of saving the space we will not write them out.

 Evidently, when the Bell states $|\Phi^+\rangle_{23}$ and $|\Phi^+\rangle_{14}$ are acquired in Alice's two
 Bell state measurements,   the probability of successful teleportation is
 \begin{equation}
\begin{array}{rl}
~~&|_{14}\langle\Phi^+|_{23}\langle\Phi^+|\Phi\rangle_w|^2(1-_{56ab}\langle\Psi'_0|P_5\otimes
I|\Psi'_0\rangle_{56ab})\\
=&\frac {|a\alpha|^2+|b\beta|^2+|c\gamma|^2+|d\delta|^2}{4}\times \frac {4}{x(\frac {1}{\alpha^2}+
 \frac {1}{\beta^2}+\frac {1}{\gamma^2}+\frac {1}{\delta ^2})(|a\alpha|^2+|b\beta|^2+|c\gamma|^2+|d\delta|^2)}\\
 =&\frac {1}{x(\frac {1}{\alpha^2}+
 \frac {1}{\beta^2}+\frac {1}{\gamma^2}+\frac {1}{\delta ^2})}.
 \end{array}
 \end{equation}
Synthesizing all  Alice's Bell state measurement cases (sixteen kinds in all), the probability of successful
teleportation in this scheme is
\begin{equation}
p=\frac {16}{x(\frac {1}{\alpha^2}+
 \frac {1}{\beta^2}+\frac {1}{\gamma^2}+\frac {1}{\delta ^2})}.
\end{equation}
Apparently, if the quantum channel is made up of the
 maximum entangled state $\frac {1}{2}(|0000\rangle+|1001\rangle+|0110\rangle+|1111\rangle)_{3456}$, i.e.
 $\alpha=\beta=\gamma=\delta=\frac {1}{2}$, we can choose  $x=1$, then  $P_5$ is zero operator. In this case
 the probabilistic teleportation  becomes usual teleportation.

 In summary, a scheme for probabilistic teleporting the unknown quantum  state of two-particle is proposed. We hope that this scheme
 will be realized by experiment.

 \acknowledgments This work was supported by Hebei Natural Science Foundation of China under Grant No:
A2004000141 and A2005000140, and  Natural Science Foundation of Hebei Normal University.


\begin{thebibliography}{s2}
\bibitem{Bennett93} C.H. Bennett et al.,  Phys. Rev. Lett.  70 (1993) 1895.
\bibitem{NielsenChuang}M.A. Nielsen, I.L. Chuang,      Quantum Computation and Quantum Information,
            Cambridge University Press, Cambridge (2000).
\bibitem{CiracZoller} J.I. Cirac, P. Zoller, Phys. Rev. Lett.  74  (1995) 4083.
\bibitem{Barenco} A. Barenco et al., Phys. Rev. Lett.  74 (1995) 4091.
\bibitem{BennenttWiesner} C.H. Bennett, S.J. Wiesner,   Phys. Rev. Lett.  69 (1992) 2881.
\bibitem{BB84} C.H. Bennett, G. Brassard,  Proc. IEEE Int. Conf. on Computers,
Systems and Signal Processing, Bangalore, India, IEEE, New York (1984).
\bibitem{Ekert} A.K. Ekert, Phys. Rev. Lett.  67  (1991) 661.
\bibitem{Bennett92} C.H. Bennett,  Phys. Rev. Lett. 68  (1992)  3121.
\bibitem{ShimizuImoto} K. Shimizu, N. Imoto, Phys. Rev. A   60 (1999) 157.
\bibitem{Beige} A. Beige  et al.,   Acta Phys. Pol. A   101 (2002) 357.
\bibitem{DengLong} F.G. Deng,  G.L. Long,   Phys. Rev. A   69 (2004) 052319.
\bibitem{YanZhang} F.L. Yan, X.Q. Zhang, Euro. Phys. J. B 41  (2004) 75.
\bibitem{Gao} T. Gao, Z. Naturforsch 59a (2004) 597.
\bibitem{BPZ}  D. Bouwneester et al., Nature,  390 (1997) 575.
\bibitem{Boschi} D. Boschi, et al.,  Phys. Rev. Lett. 80  (1998) 1121.
\bibitem{Furusawa} A. Furusawa, et al.,  Science  282  (1998) 706.
\bibitem{GorbachevTrubilko} V.N. Gorbachev, A.I. Trubilko,  JETP  91 (2000) 894.
\bibitem{LuGuo} H. Lu, G.C. Guo, Phys. Lett.  A 276  (2000) 209.
\bibitem{Bandyopadhyay}  S. Bandyopadhyay,  Phys. Rev.  A 62  (2000)  012308.
\bibitem{Rigolin}  G. Rigolin,   Phys. Rev. A 71  (2005) 032303.
\bibitem{ShiGuo}  B.S. Shi, Y.K. Jiang, G.C. Guo,  Phys. Lett.   A 268  (2000) 161.
\bibitem{ZengLong} B. Zeng, X.S. Liu, Y.S. Li,  G.L. Long, Commun. Theor. Phys. 38  (2002) 537.
\bibitem{MorHorodecki} T. Mor,   P. Horodecki, quant-ph/9906039.
\bibitem{LiuGuo} J.M. Liu,  G.C. Guo,  Chin. Phys. Lett.   19 (2002)  456.
\bibitem{DaiLi}  H.Y. Dai, C.Z. Li, Chin. Phys. Lett.  20 (2003) 1196.
\bibitem{YanTanYang}  F.L. Yan, H.G. Tan,  L.G. Yang, Commun. Theor. Phys. 37 (2002) 649.
\bibitem{YanWang}  F.L. Yan, D. Wang, Phys. Lett. A  316  (2003) 297.
\bibitem{YanYang}  F.L. Yan,   L.G. Yang, IL Nuovo Cimento   118B  (2003) 79.
\bibitem{YanBai}  F.L. Yan, Y.K. Bai, Commun. Theor. Phys.  40  (2003) 273.
\bibitem{GaoYanWang}  T. Gao, F.L. Yan, Z.X. Wang, Quantum Inform. Comput.  4  (2004) 186.
\bibitem{GaoWangYanChinPhysLett}  T. Gao, Z.X. Wang, F.L. Yan, Chin. Phys. Lett. 20 (2003) 2094.
\bibitem{GaoYanWangChinesePhysics}  T. Gao, F.L. Yan, Z.X. Wang, Chin. Phys.  14 (2005) 893.
\bibitem{Gao2}  T. Gao,  Commun. Theor. Phys. 42  (2004) 223.
\bibitem{Helstrom} C.W. Helstrom, Quantum Dectection and Estimation Theory, Academic Press, New York (1976).
\end{thebibliography}
\end{document}